\documentclass[fleqn,10pt]{wlscirep}

\title{ Magnetoresistance and robust resistivity plateau in MoAs$_2$}

\author[1]{Jialu Wang$^{1}$, Lin Li$^{1}$, Wei You$^{1}$, Tingting Wang$^{1}$, Chao Cao}
\author[1,*]{Jianhui Dai$^*$, \& Yuke Li}
\affil[1]{Department of Physics and Hangzhou Key Laboratory of Quantum Matter, Hangzhou Normal University, Hangzhou 310036, China}
\affil[*]{daijh@hznu.edu.cn,yklee@hznu.edu.cn}

\begin{abstract}

We have grown the MoAs$_2$ single crystal which crystallizes in a
monoclinic structure with C2/m space group. Transport
measurements show that MoAs$_2$ displays a metallic behavior at zero
field and undergoes a metal-to-semiconductor crossover at low temperatures
when the applied magnetic field is over 5 T. A robust resistivity
plateau appears below 18 K and persists for the field up to 9 T. A
large positive magnetoresistance (MR), reaching about 2600\% at 2 K and 9 T, is observed when the field is perpendicular to the current.
The MR becomes negative below 40 K when the field is rotated to be parallel to the current. The Hall resistivity
shows the non-linear field-dependence below 70 K. The analysis using two-band model indicates a compensated electron-hole carrier density at low temperatures. A combination of the breakdown of Kohler's rule, the abnormal drop and the cross point in Hall data implies that a possible Lifshitz transition has occurred between 30 K and 60 K, likely driving the compensated electron-hole density, the large MR as well as the metal-semiconductor transition in MoAs$_2$. Our results indicate that the family of centrosymmetric transition-metal
dipnictides has rich transport behavior which can in general exhibit variable metallic and topological features.

\end{abstract}

\begin{document}

\flushbottom
\maketitle
\thispagestyle{empty}

\section*{Introduction}

Three-dimensional (3D) topological quantum materials, including
topological insulators (TIs)\cite{fu2008superconducting}, topological Dirac semimetals
(DSMs)\cite{NaBi1,NaBi2,NaBi3,CdAs2,CdAs3} and Weyl semimetals
(WSMs)\cite{TaAs,TaAs2,TaAs3,TaAs4,NbAs,NbAs2,NbP,TaP}, have been
discovered and intensively investigated recently. These materials
exhibit a variety of interesting physical properties, owing to their
unique electronic structures and spin textures\cite{NaBi1,CdAs3,TaAs2,TaAs4,Weng}, and thus show a
broad application potential. An ideal TI usually exhibits a clear
resistivity plateau at low temperatures due to the robust surface
states\cite{Ren,Jiashuang,Wolgast13,Kim13}.  In an ideal WSM, the
Weyl fermions disperse linearly all the way across the Weyl nodes
which appear in pairs with opposite chiralities by breaking time reversal symmetry (TRS) or
inversion symmetry\cite{Weng,WanX,Xu}. The relativistic electronic
dispersion and chirality-based topological property result in
various semi-metallic transport properties and produce a number of
novel phenomena such as the anomalous Hall effect\cite{TaAs} and the
Fermi arcs\cite{TaAs3}.

%In an ideal 3D TI where the bulk states
%are completely gapped out near the Fermi level, a resistivity
%plateau can clearly develop in the low temperature regime because
%the only participating surface states are robust to disorders as
%protected by time-reversal symmetry

Recently, the WSMs have been theoretically predicted and
experimentally discovered in a family of transition-metal pnictides
represented by TaAs family\cite{TaAs,TaAs2,TaAs3,TaAs4,NbAs,NbAs2,NbP,TaP,Weng}.
Consequently, dozens of topological semimetals showing exotic
physical properties were reported and studied in detail, such as
ZrSiS(Te)\cite{ZrSiS,ZrSiTe}, MoTe$_2$\cite{MoTe,MoTe2} and
WTe$_2$\cite{WTe2Cava}, in addition to the TaAs family. All these materials
are non-centrosymmetric in the crystal structure and are further
classified into two types of Weyl fermions: with or without the
Lorentz symmetry in their energy-momentum dispersions. More
recently, a new family of topological semimetals (TSMs) which
crystallizes in the centro-symmetry monoclinic structure, the
transition-metal dipnictides $XPn_2$ ($X$=Ta, Nb, $Pn$=P, As, Sb) as
represented by TaSb$_2$, have been discovered.\cite{TaSbli,TLXia,ZFang}

The reported transport properties of the centrosymmetric $XPn_2$ exhibit the extremely large
magnetoresistance (MR) and ultrahigh electronic mobility in common\cite{YKLuo,Sjiang,NNI,ZAXu}.
Meanwhile they also exhibit the material's-dependent low-temperatures
resistivity plateau and negative MR\cite{TaSbli,YKLuo,KWang}. While the explanations of these
features are still complicated, they are likely related to the
disentangled bulk electron/hole bands and topological non-trivial
surface states as revealed in the electronic band structure
calculations\cite{Cao:prb16}. On the other hand, the coexistence of
bulk and surface states is appealing for materials applications. We
are thus motivated to search for other candidates in this family
with potentially more robust surface states and better metallicity.
%Apart from the perspective of fundamental physics, the
%magnetotransport properties such as extremely magnetoresistance
%(MR), negative MR and ultrahigh electronic mobility
%Therefore, experimentally finding the new materials with particular
%magnetotransport properties are of highly desirable.

In this paper, we report a new member of this
family, MoAs$_2$, which also crystallizes in a monoclinic structure.
Our experimental results indicate that MoAs$_2$ undergoes a
metal-to-semiconductor crossover under the applied magnetic field up
to 5 T. In particular, a very clear resistivity plateau is observed
below 18 K even in the absence of the magnetic field. The plateau
feature is robust against the applied magnetic fields up to 9 T. The
MR is relatively large as the applied magnetic field is perpendicular to the
current, reaching about 2600\% at 2 K and 9 T, but violates the Kohler's rule below 70 K.
When the applied field is rotated from perpendicular to parallel to the current
direction, the MR drops rapidly and finally becomes negative. Hall resistivity shows the non-linear field-dependence below 70 K
and changes from positive to negative at around 40 K. The two-band model fit yields the abnormal drop in hole-mobility below 30 K and a strange cross point in carrier density at around 60 K. All our experimental results indicate a possible Lifshitz transition occurring between 30 K and 60 K, which is likely a direct factor to drive the compensated electron-hole carrier density, the large MR as well as the metal-semiconductor-like transition in MoAs$_2$.

\section*{Results}

Fig. 1 shows the detailed information of MoAs$_2$ crystal structure.
In Fig. 1a, MoAs$_2$ crystallizes in a monoclinic structure with C2/m space group, which is common
to the $XPn_2$ family. Fig. 1b shows the (00l) peaks of MoAs$_2$ single crystal X-ray diffraction,
implying that the crystal surface is
normal to the c-axis. The b-axis is perpendicular to the ac-plane,
and parallel to the current direction. The left inset of Fig. 1b shows a
picture of a large polyhedral MoAs$_2$ crystal with millimeter
dimension. The polyhedral crystal with rectangle-shape is consistent
with the monoclinic structure. The right inset shows the X-ray rocking curve with a very small half-high-width, implying the high quality crystal.
Fig. 1c exhibits the Rietveld refinement profile and its structure parameters in MoAs$_2$ at room
temperature (more details of crystal structure parameters can be found in Table S1 in Supplementary Information). Through the Rietveld structural analysis, the refined lattice parameters are extracted to be $a=$ 9.064(7) {\AA}, $b=$ 3.298(7) {\AA}, $c=$  7.718(3) {\AA}, and $\beta =$119.37(1)$^\circ$ as reported in the previous literatures\cite{MoAs,MoAs2}. Fig. 1d shows the EDX data with the Mo and As contents of 33.3\% and 66.7\%, respectively, consistent with the ratio of Mo and As.

Fig. 2 displays the evolution of the magneto-resistivity as a
function of temperature in MoAs$_2$ down to 2 K with the magnetic field $\textbf{B} \parallel c \bot \textbf{I} $. As shown in figure 2a, the value of $\rho_{xx}$ at 300 K is about 163 $\mu \Omega$ cm at \textbf{B} = 0 T, comparable to that of high quality WTe$_2$\cite{WTe2Cava} and
MoTe$_2$\cite{MoTe}, the potential candidates of the type-II WSM.
While, $\rho_{xx}$ at 2 K falls to 0.29 $\mu \Omega$ cm, yielding a large
value of residual resistivity ratio (RRR) $=$ 562. This confirms the high quality of the crystal. The
extraordinarily low residual resistivity at 2 K was previously found
in Cd$_3$As$_2$\cite{Cd2As3}, ZrSiS\cite{ZrSiS3}
and high purity Bi\cite{Bi} with very large RRR.

The zero field resistivity $\rho_{xx}$ exhibits highly metallic
behavior. Upon decreasing temperature, it decreases almost linearly
in the high temperature regime. Below 30 K, however, $\rho_{xx}$ deviates severely from the Fermi liquid behavior.
Considering that this contribution to resistivity involves the electron-phonon(e-ph) interaction according to the Bloch-Gruneisen theory\cite{twoband}. We employed the formula $\rho = \rho_0 + a\emph{T}^2 + b\emph{T}^5$ to fit the low temperature resistivity as shown in inset of Fig. 2a, where the $\rho_0$ is the residual resistivity at $T$ = 0 K, the terms of \emph{T}$^2$ and \emph{T}$^5$ represent the contributions of the e-e and e-ph scatter, respectively. It is found that resistivity does not exactly match this fit. Instead it remains almost constant below 20 K, displaying a clear resistivity plateau.

An applied magnetic field does hardly change the resistivity in the high temperature regime as compared with that of
the zero field. A significant change in $\rho_{xx}$ appears around $T_m =$ 40K,
where the resistivity starts to saturate for small field or
crossovers to the semiconductor behavior for large field. After the
crossover the resistivity is soon saturated, developing a robust
resistivity plateau in the low temperature regime below $T_i =$ 18 K, defined as
the onset temperature of the plateau where $\partial^2{\rho}(T)/\partial{T^2}=0$, as clearly observed in the Fig. 2b. The crossover behavior becomes prominent with increasing magnetic field up to 9 T while the
crossover temperature does not change too much.

A large MR ($=(\rho_{xx}(B) - \rho_{xx}(0))/\rho_{xx}(0)$) is observed, reaching 2600$\%$ at 2 K and 9 T in figure 2c. Although the MR in MoAs$_2$ is one or two orders of magnitude smaller than that in other semimetal compounds, it is still far larger than the value in some ferrimagnetic materials\cite{LaSrMnO}. According to the Kohler's rule\cite{twoband,WTe2Kohler} (MR $\propto (B/\rho{(T,0)})^m$), all the MR at various temperatures can be scaled onto a single line with a constant $m=1, 2$. In figure 2c, the fit of MR curve at 2 K yields $m = 1.7$. A careful analysis to the MR across the metal-semiconductor-like transition regime found that the $m$, as shown in inset of figure 2c, displays a minimum value around 40 K for $m = 1.5$, but then increases slightly to about 1.7 below 10 K. The small $m$ value substantially violates the classically field-square dependence of the MR in the semiconductor-like regime. Noted that the exponent $m$ ($\sim$ 1.5) has been observed in under-doped cuprates and organic conductor which both are widely considered as non-Fermi liquid system\cite{organic,cuprates}. The results, as plotted in figure 2d, display that the MR below 70 K obviously deviates from the Kohler's rule, indicating that the sample is a multi-band system possessing the variable carrier concentrations or the mobility ratio of electron to hole when temperature goes down. Therefore, we suggest that the conventional e-e and e-ph interactions may be hard to explain the appearance of resistivity plateau in MoAs$_2$, implying that this intrinsic plateau is closely related to the topological nature as observed in many semimetal materials\cite{LaSb,TaSbli}.

However, in these semimetal compounds, saturated resistivity
plateaus at much lower temperatures are naturally expected. The
microscopic origin of this phenomenon remains debated, depending on
the expanded energy scale around which the plateau feature sets in.
In the candidates of TIs, SmB$_6$ \cite{Wolgast13,Kim13} and
LaSb\cite{LaSb,FuL}, the resistivity plateau is well-attributed to
the topological non-trivial surface states which are robust against
disorders. It is interesting that a clear resistivity plateau was
also observed in TaSb$_2$, a candidate of TSM \cite{TaSbli}.
Although bulk excitations are involved in this material, the
resistivity plateau is plausible due to the surface states which are
topological non-trivial in the sense of doped weak
TIs\cite{Cao:prb16}. It is remarkable that the resistivity plateau
of MoAs$_2$ sets in at $T_i =$ 18 K, about four times of that of
SmB$_6$ ($T =$ 5 K )\cite{Kim13}.

%FIELD DEPENDANCE%%%%%%%%%%%%%%%%%%%%%%%%%%%%%%%%%%%%%%%%%%%%%%%%%%%%%%%%%%%%%%%%%%%%%%%%%%%%%%%%%%%%%%%%%%%%%%%%%%%%%%%%%%%%%%%%%%%%%%%%%%%%%%%%%%%%%%%%%%%%%%%%%%%%%%%%%%%%%%%%%%%%%%%%%%%%%%

Now we turn to the MR of MoAs$_2$ as shown in Fig. 3.  All data here are
displayed without a symmetrizing process. In Figure 3a, a large MR at low temperatures is observed when the field
is perpendicular to the current direction.
%This value is two or three orders of magnitude smaller than that
%of other semi-metal materials with a comparable RRR.
The value of MR does not change significantly until 20 K where it
decreases sharply in consistent with the robust resistivity plateau
which persists up to 9 T. At fixed temperatures, the MR increases
quadratically for low field and almost linear for larger field
without saturation, similar to the previously-known semimetallic
materials including TaAs(P)\cite{TaAs,TaP}, NbAs(P)\cite{TaAs,TaP}
and WTe$_2$\cite{WTe2Cava}. Further increasing temperature, its value
approaches to 3\% at 100 K, and is less than 1\% above 200 K at 9 T.
This feature is in contrast to the most of semimetals with a relatively
large MR even at room temperatures\cite{TaP,ZrSiS}.
%because electrons or holes mobility of MoAs$_2$ is rather lower than that of other semimetals at high temperatures, consistent with our Hall data displayed below\cite{note4}.

As the magnetic field is applied along the current direction,
${\textbf B}\parallel{\textbf I}$, the MR becomes negative at low
temperatures shown in figure 3b. These MR curves
are overall axial-symmetric around B $=$ 0 T, with only slight
noises due to the extraordinarily low resistivity. The absolute
value of MR is much smaller than that in the case of $ {\textbf
B}\bot {\textbf I}$.
%shown previously.
%Furthermore, the
%MR in the whole temperature regime lowers below 100\%, far more
%smaller than that in the case of $ {\textbf B}\bot {\textbf I}$
%shown previously.
At 2 K, the MR first increases until about 2 T, then decreases
monotonously with the field and changes a sign from positive to
negative as B $\sim $ 6.5 T, reaching about 20\% at 9 T. The
negative MR decreases gradually upon heating and disappears above 40
K, beyond which the MR turns back to positive.

Fig. 3c shows the field dependence of MR at various angles $\theta
$ of the magnetic field with respect to the current direction at 2
K. By rotating $\theta = 90^\circ\rightarrow 0^\circ$, the MR drops
quickly and shows quadratically field-dependence. The negative MR is
further illustrated in figure 3d when $\theta$
approaches to $0^\circ$.  At B $=$ 9 T, the MR is still negative for
$\theta \leq 4^\circ$. When $\theta$ is larger than 4$^\circ$, the
MR recovers positive and increases with $\theta$. The negative MR at
$\theta $ = 0$^\circ$ is limited in a narrow window over $\sim$ 6.5 T.
The phenomenon of negative MR has been observed in a number of
metallic compounds with high mobility of charge carriers. The
interpretation of this phenomenon remains debated, too, given the
fact that it may appear in both topological trivial and non-trivial
materials. A crucial issue behind the negative MR is the current
jetting effect due to the field-induced anisotropy. This effect is
usually elusive but has been recently suggested as a main cause of
negative MR in TaP compound\cite{TaP,current}. In the TaSb$_2$
samples, however, the current jetting effect was shown to play a
minor influence on the negative MR by using different contact
configurations\cite{TaSbli}. In the ideal WSMs, the negative MR
could be best-understood due to the chiral anomaly. The similar
topological interpretation applies to those with ill-defined Weyl
points or Dirac points if they appear in pairs and separate in
momentum space\cite{Weylpoint,illweylpoint}. In the present compound
MoAs$_2$, the topological interpretation seems to be consistent with the
robust resistivity plateau discussed previously.

%The similar $\theta$ window was observed in Na$_3$Bi and TaP
%compounds\cite{TaP,Xiongjun}. However, the negative MR decreases
%monotonically with fields reaching about -20\% at B $=$ 9, and tends
%to unsaturate within the present magnetic field. This remarkable
%feature is also observed in the recent reported semimetal compounds
%such as TaSb2, NbP, NbAs2. As a comparison, the negative MR in other
%semimetal materials including Na$_3$Bi\cite{Xiongjun},
%Cd$_3$As$_2$\cite{CLi,CZhang,ShenSQ}, TaAs\cite{GFChen},
%NbP\cite{Zwang}, is limited not only in a narrow window of $\theta$,
%but also in a window of B, namely, the negative MR will saturate and
%return to positive for larger $B$ even for $\theta = 0^{\circ}$.

%According to the chiral anomaly\cite{ABL,ABL2}, the MR should remain negative as long as ${\textbf B}||{\textbf
%I}$. If imperfect alignment of the
%magnetic field and the current in samples can be fully excluded, the
%limited window in B should be due to the
%disorder-induced weak localization. So the unsaturated negative MR
%observed here implies the consistency with the chiral
%anomaly interpretation and the high quality of the measured
%samples.

Fig. 4. shows the band structure with inclusion of spin-orbit coupling
(SOC). The inversive bands from X$_1$ to Y lead
to a Dirac cone in the absence of the SOC, while it opens a gap when
the SOC is taken into account, resulting in two disentangled and
nearly compensated electron/hole bands. We can calculate the
topological index for each of these bands according to the
parity-check method\cite{Fu}. Different to TaSb$_2$ or other
previously-known $XPn_2$ compounds, the indices of the
partially-filled bands of MoAs$_2$ are topologically
strong\cite{note}. The result indicates that the surface states in
MoAs$_2$ are more robust than those in TaSb$_2$\cite{TaSbli}, consistent with the
observed resistivity plateau in our present sample.

%Halleffect and QSH%%%%%%%%%%%%%%%%%%%%%%%%%%%%%%%%%%%%%%%%%%%%%%%%%%%%%%%%%%%%%%%%%%%%%%%%%%%%%%%%%%%%%%%%%%%%%%%%%%%%%%%%%%%%%%%%%%%%%%%%%%%%%%%%%%%%%%%%%%%%%%%%%%%%%%%%%%%%%%%%%%%%%%%%%%%%%%

Fig. 5a maps the magnetic field dependence of the Hall resistivity
for MoAs$_2$ at various temperatures, with the magnetic field
ranging from 0 to 9 T. The field dependence of $\rho_{xy}$(B) is
almost linear with positive slope at high temperature. It starts to
bend strongly below 70 K. For T$ \leq$ 40 K, $\rho_{xy}$(B) shows a
pronounced sign reversal from positive to negative, and finally
recovers the nearly linear behavior, implying a multi-band system in MoAs$_2$.
Fig. 5b shows the temperature dependence of the Hall coefficient at 1 T, 4 T, and 9 T. As
temperature decreases, R$_H$ firstly increases from 300 K, and then
undergoes a sharp drop at around 70 K. R$_H$ changes its sign from
positive to negative at around 40 K, implying a partial compensation
between the hole-type and electron-type carriers at low temperatures. The sign-change in R$_H$ is usually (though not always) an important indication of multiple bands, so the feature of the sign-changed R$_H$ confirms further the multi-band effect in MoAs$_2$. The sign-changed R$_H$ together with the increase of electron-mobility and -density as shown in figure 6 implies a possible temperature-induced Lifshitz transition occurring at around temperature, resulting in a drop R$_H$ below 70 K associated with the decrease of hole pockets.

%Using a single formula $n = 1/eR_H(0)$, where $R_H(0)$ is the zero temperature limit of $R_H(B)$, the electron-type carrier
%concentration is estimated to be $n_e$ = 8 $\times 10^{21}$ at 9 T.
%The calculated carrier mobility is about 2.69$\times 10^3$
%$cm^2V^{-1}s^{-1}$ at 2 K.

%Hence MoAs$_2$ has a relatively high
%carrier density of electron-type and a low carrier mobility compared
%to the typical semimetal compounds\cite{LaSb}.
%The strong nonlinear behavior
%in $\rho_{xy}$ and the sign changed $R_H$ in several semimetals
%\cite{NbP,TaAs2,Cava} have been suggested to be a signature of the
%electron-hole compensation.

We carefully analyze the Hall conductivity according to the two-band model\cite{twoband}. The results of the densities and mobilities of both electrons and holes as a function of the temperature are shown in Fig.6. In this model, the Hall conductivity tensor is given by:

\begin{equation}
\sigma_{xy} = -\frac{\rho_{xy}}{{\rho_{xx}}^2+{\rho_{xy}}^2}
\\
\sigma_{xy} = eB[\frac{n_h{\mu_h}^2}{1+{(\mu_hB)}^2}-\frac{n_e{\mu_e}^2}{1+{(\mu_eB)}^2}]
\end{equation}

Here, $n_e(n_h)$ and $\mu_e(\mu_h)$ are electrons (holes) carrier densities and mobilities, respectively. The result gives a excellent fit below 100 K and the obtained $n_{e,h}$ and $\mu_{e,h}$ as a function of temperature are shown in figure 6c and 6d. Above 70 K, the hole-type carrier dominates, consistent with the positive Hall coefficients as shown in figure 5b. With lowering temperature, the electron-type carrier density increases sharply and becomes dominant, resulting in an observed cross point at about 60 K which is very closed to the characteristic temperature ($T_m$) with the sign-changed R$_H$. As further cooling down temperature, the $n_{e,h}$ increases slowly, and then becomes almost equivalent and saturated. The calculated density of electrons and holes at 2 K is 3.95 $\times 10^{20}$ cm$^{-3}$ and 4.13 $\times 10^{20}$ cm$^{-3}$, respectively, implying a perfect compensation of semimetal in MoAs$_2$. In figure 6d, the mobility, $\mu_{e,h}$, as a function of temperature increases slightly and displays more or less overlap above 30 K, followed by a huge divergency that $\mu_{h}$ drops suddenly and $\mu_{e}$ increases sharply. Noted that the metal-semiconductor transition occurs at around 40 K, below which the MR starts to increase dramatically. A combination of the drop in mobility, the cross point in $n_{e,h}$ and the sign-changed R$_H$ points to a possible electronic structure change at around 30-60 K, named a Lifshitz transition, directly driving the compensation of electron-hole carriers density at low temperature and the extremely large MR as well. The similar behavior has been investigated in the discovered WTe$_2$ and MoTe$_2$ semimetals by the measurement of ARPES\cite{WTe2Lifshize,WTe2ARPES}, Ultrafast transient reflectivity\cite{UFAR} and Hall effect\cite{MoTe2b,WTe2YKLuo}.

Therefore, the possible Lifshitz transition can be suggested in MoAs$_2$, which directly derives a compensated electron-hole carriers density, the large MR as well as the metal-semiconductor-like transition. More experiments in MoAs$_2$ such as ARPES need be investigated in future to clarify this phase transition, since it is likely to make out the origin of those transport properties and its topological properties.

\section*{Discussion}

Presently, the large MR has been discovered and investigated in magnetic multilayers\cite{Multilayer} and semiconductor 2D-electronic systems\cite{Add,Add1,Add2,Add3}, in which the rearranged magnetic moments and electron-electron interaction likely play a key role. However, the large unsaturated positive MR recently discovered in some nonmagnetic semimetals\cite{NaBi3,Cd2As3,TaAs,TaSbli} seems to be unique and puzzled.
The possible mechanisms can be mainly ascribed to the several factors: (1) the linear energy dispersion of Dirac fermions leading to the quantum limit\cite{QuantumL}; (2) the compensated electron-hole carrier density\cite{WTe2Cava,TaAs2}; (3) the turn on temperature behavior following the Kohler's rule\cite{WTe2Kohler}; (4) other reasons such as protection mechanism\cite{Cd2As3}. In MoAs$_2$, the positive MR shows the almost quadratic field-dependent not linear at low magnetic fields, implying that the quantum limit should not dominate the transport. MoAs$_2$ also displays the turn on temperature behavior as B $\geq$ 5 T, analogous to the other most of semimetals, but its MR clearly violates the Kohler's rule below 70 K. On the other hand, the sign-changed R$_H$ and the almost equal amount of electron and hole density at low temperatures have been observed in the present sample. We thus suggest that the large MR can be attributed to the electron-hole compensation, similar to WTe$_2$\cite{WTe2Cava}.

Unlike the high mobility of the two-dimensional electron gas (GaAs/AlGaAs heterostructure)\cite{Add5},
the negative MR associated with the chiral anomaly has been viewed as a key transport signature in semimetal compounds, such as TaSb$_2$\cite{TaSbli}, TaAs$_2$\cite{YKLuo}, and Cd$_2$As$_3$\cite{NegativeCd2As3}. When the magnetic field is parallel to the electric current, the chiral anomaly induces the non-conservation of the fermions number with a given chirality, resulting in the negative MR. Recently, an alternative scenario for the negative MR has been suggested, called as the current jetting effect which is associated with an inhomogeneous current distribution inside the sample in a magnetic field\cite{current,TaP}. We thus perform the different contact configurations on resistivity measurements (see the figure S1 in Supplementary Information). Consequently, the negative MR is robust and shows weakly dependent on the different contact configurations. Therefore, we believed that the current jetting effect may play a minor role in the negative MR in our present sample, as far as in the XPn$_2$ system\cite{TaSbli,YKLuo}.

In summary, we reported an iso-structural MoAs$_2$ compound which
crystallizes in the monoclinic structure with a centrosymmetric
space group C2/m. It shows large positive MR and undergoes a metal-semiconductor crossover under 9 T, but clearly violates the Kohler's rule below 70 K. At low temperatures, a robust resistivity plateau with T$_i$ = 18 K is observed up to 9 T, which is likely caused by the topological non-trivial surface states whose existence should be further examined by the ARPES experiment. When B $\parallel$ I, the negative MR is observed regardless of the different contact configurations, implying a minor current jetting effect in our sample. On the other hand, The analysis of two-band model in Hall resistivity indicates an almost equal amount of electrons and holes at low temperature, consistent with the linear Hall resistivity. A combination of the drop in hole-mobility, the cross point in density and the sign-changed $R_H$ as a function of temperature together with the breakdown of Kohler's rule suggests a possible Lifshitz transition occurring between 30 K and 60 K, likely driving the electron-hole compensation, the large MR as well as the metal-semiconductor-like transition in MoAs$_2$.

\section*{Methods}

Very high quality single crystals of monoclinic MoAs$_2$ were grown
through chemical vapor transport reaction using iodine as transport agent.
Polycrystalline samples of MoAs$_2$ were first synthesized by
solid state reaction using high purified Tantalum powders and
Antimony powders in a sealed quartz tube at 973 K for three days.
Subsequently, the final powders together with a transport
agent iodine concentration of 10 mg$/cm^3$ were ground thoroughly, and then were sealed in a quartz tube with a low Vacuum-Pressure of $\leq$ 10$^{-3}$ Pa. The single crystals
MoAs$_2$ were grown in a horizontal tube furnace with a temperature gradient of 120 $^{\circ}{\rm C}$ between 1120 $^{\circ}{\rm C}$ -1000 $^{\circ}{\rm C}$ for 1-2 weeks. The high quality single crystals with apparent monoclinic shape were picked from the resultant.

X-ray diffraction patterns were
obtained using a D/Max-rA diffractometer with CuK$_{\alpha}$
radiation and a graphite monochromator at the room temperature. The
single crystal X-ray diffraction determines the crystal grown
orientation. The composition of the crystals was obtained by energy
dispersive X-ray (EDX) spectroscopy. No iodine impurity can be
detected in these single crystals. The stoichiometric ratio is fairly homogenous.
The (magneto)resistivity and Hall coefficient measurements were performed using the standard
four-terminal method in which the current is parallel to the b-axis. Ohmic contacts were carefully prepared on the crystal with a Hall-bar geometry. Silver wires were attached on the surface of crystals with silver paste. The low contact resistance was obtained after annealing at 573 K for an hour. The physical properties were performed in a commercial Quantum Design PPMS-9 system with a torque insert with a temperatures range from 2 to 300 K and the magnetic fields up to 9 T. The density functional theory calculations of MoAs$_2$ were performed by
employing the plane wave basis projector augmented wave (PAW) method
implemented in the Vienna ab-initio Simulation Package\cite{vasp1,vasp2}. The electronic structure calculations
were then performed using crystal structures with optimized lattice
constants and internal atomic parameters.

\section*{Acknowledgements}

This research was supported in part by the NSF of China (under
Grants No. 11274006 and No. 11274084) and the National Basic
Research Program (under Grant No. 2014CB648400). Yu-Ke Li was supported by an open
program from Wuhan National High Magnetic Field Center (2016KF03). We would like to thank Yi Liu, Jianhua Du and Qunlin Ye for technique assistances. The authors are grateful to Hangdong Wang, Shiyan Li for stimulating discussions.

\section*{Author contributions}

Y. Li designed the research. J. Wang synthesized the samples. J. Wang
and Y. Li performed most measurements. L. Li, W. You and T. Wang assisted
the measurements. C. Cao performed the band structure calculations. J. Dai, and Y. Li,
discussed the data, interpreted the results, and wrote the paper.

\section*{Additional information}
Competing financial interests: The authors declare no competing financial interests.
Correspondence and requests for materials should be addressed to
Yuke Li(email: yklee@hznu.edu.cn).

\begin{figure}[ht]
\centering
\includegraphics[width=16cm]{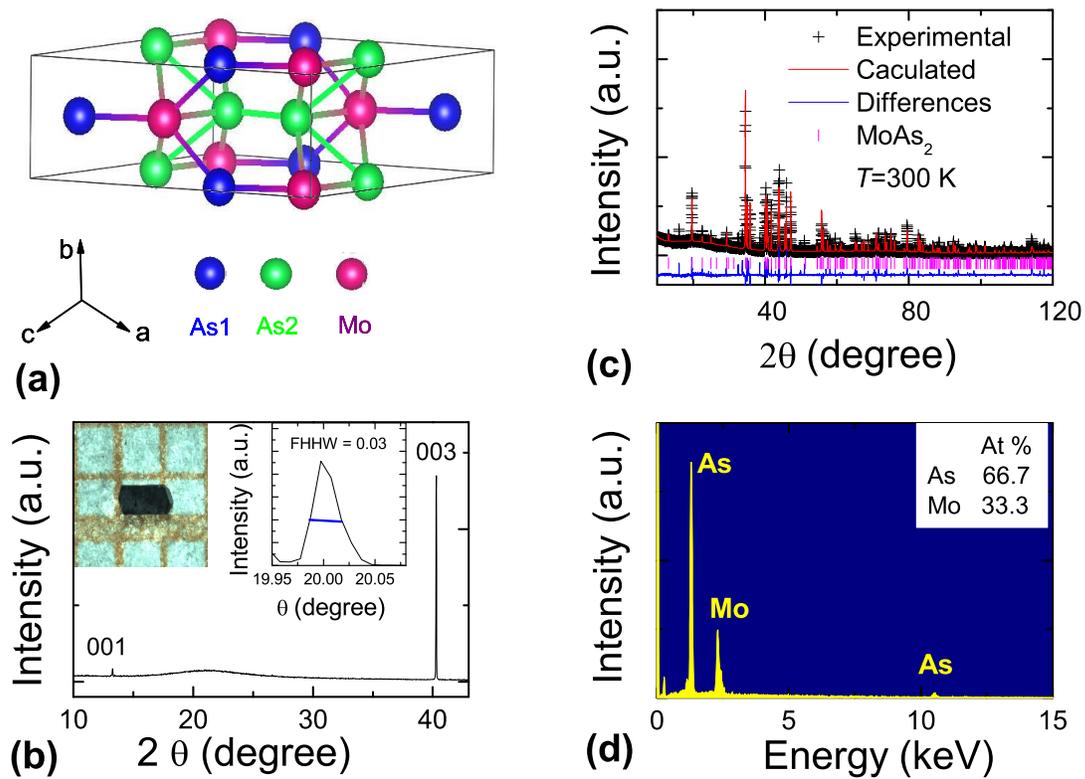}
\caption{\label{Fig.1}{\bf Crystal Structure.} (a) The crystal structure of MoAs$_2$. (b) the (00l) X-ray diffraction
patterns of single crystal. Left inset: picture of a typical MoAs$_2$ single crystal. Right inset: the rocking curve with a very small half-high-width. (c) The Rietveld refinement profile for MoAs$_2$ sample at room temperature. (d) EDX spectroscopy at room temperature.}
\label{fig:stream}
\end{figure}

\begin{figure}
\includegraphics[width=16cm]{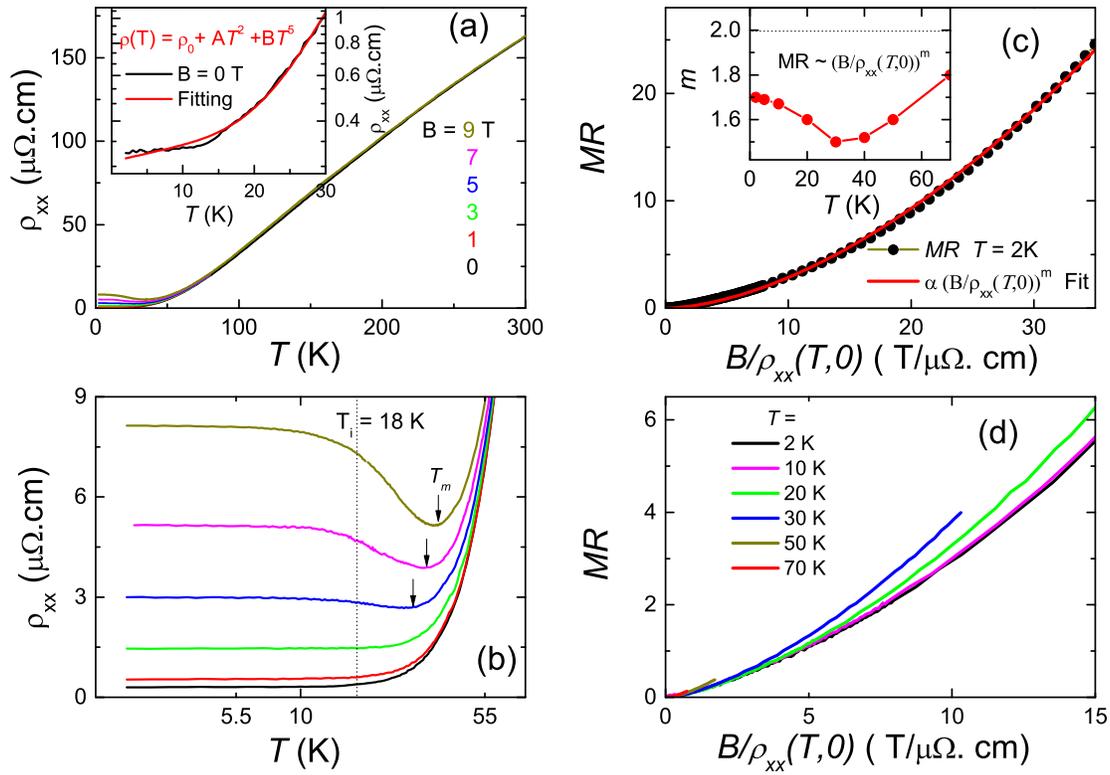}
\caption{{\bf Temperature dependence of resistivity for MoAs$_2$ and the Kohler's rule below 70 K.} (a) Resistivity of MoAs$_2$ as a function of temperature in several magnetic fields ($B = 1, 3, 5, 7, 9$ T.)
perpendicular to the current. Inset shows a deviated Fermi liquid behavior below 30 K at B = 0. The (b) shows a clear plateau resistivity at low temperature. (c) MR vs. B/$\rho_0$ at 2 K. The red line represents the fitting curve. The inset shows the fitting $m$ values at different temperatures. (d) The breakdown of Kohler's rule below 70 K. }
\end{figure}

\begin{figure}
\includegraphics[angle=0,width=16cm,clip]{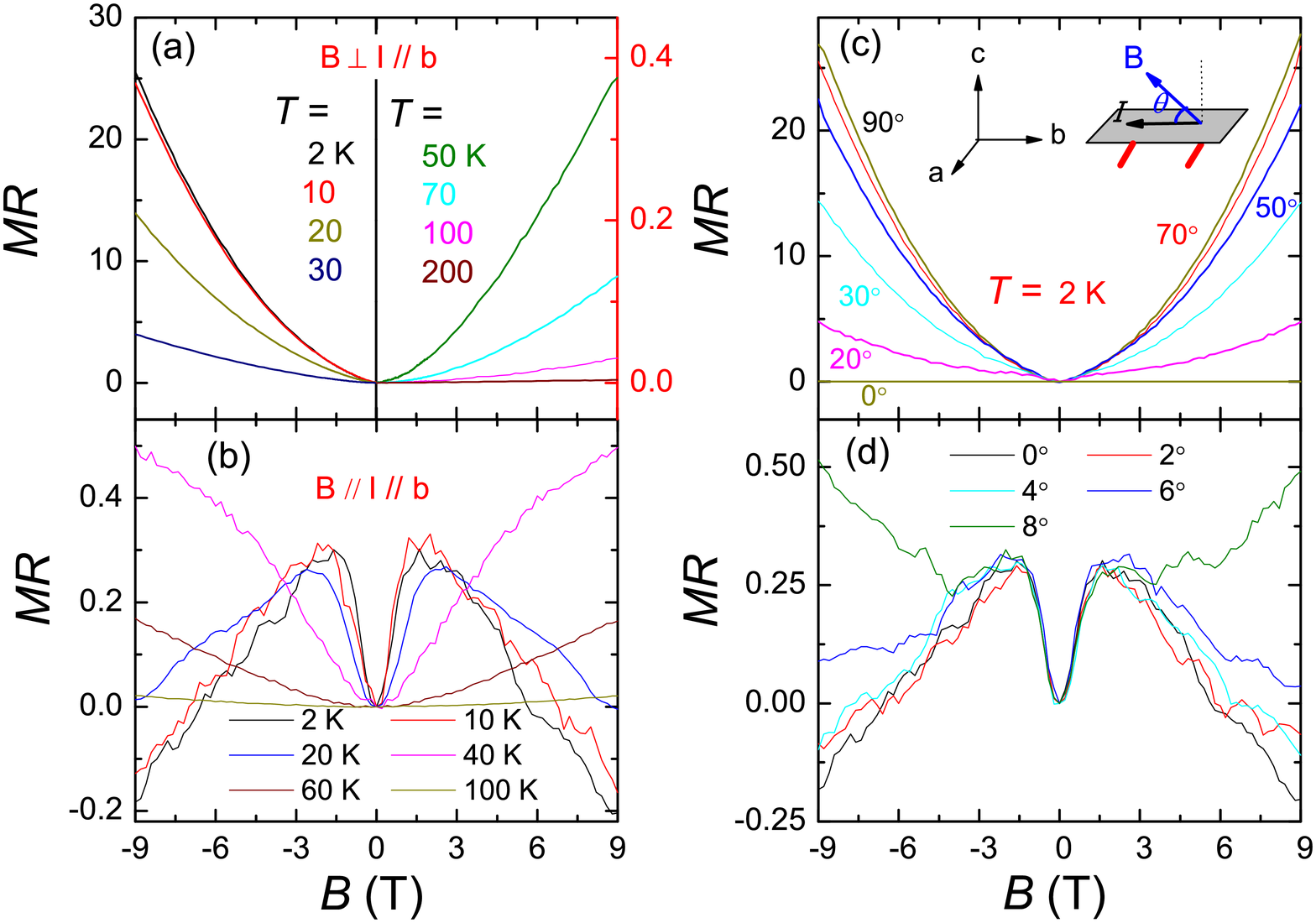}
\caption{{\bf Magnetic field dependence of MR in
MoAs$_2$ single crystal.} (a), Magnetoresistance (MR $=(\rho_{xx}(H)-\rho_{xx}(0))/\rho_{xx}(0)$) versus
magnetic fields along the c-axis at different temperatures as $
\textbf{B}$ $\perp$ \textbf{I} $\|$ b. (b), MR vs. fields for
$\textbf{B} \|\textbf{I} \|$ b. (c), MR plotted as a function of magnetic fields at
different angles between $\textbf{B}$ and $\textbf{I}$. (d), The large and unsaturated negative MR emerges in a narrow
window of angle around $\theta = 0^\circ$.}
\end{figure}

\begin{figure}
\includegraphics[angle=270,width=16cm,clip]{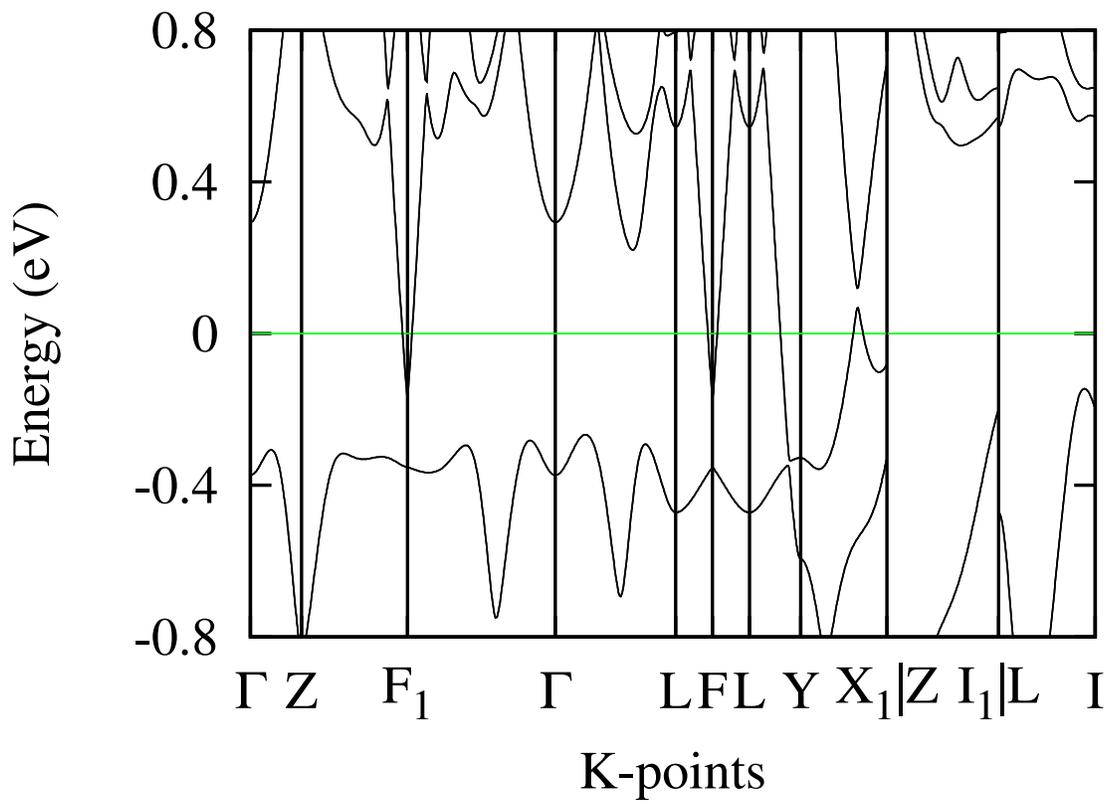}
\caption{ Band structure of MoAs$_2$ with SOC. }
\end{figure}

\begin{figure}
\includegraphics[angle=0,width=16cm,clip]{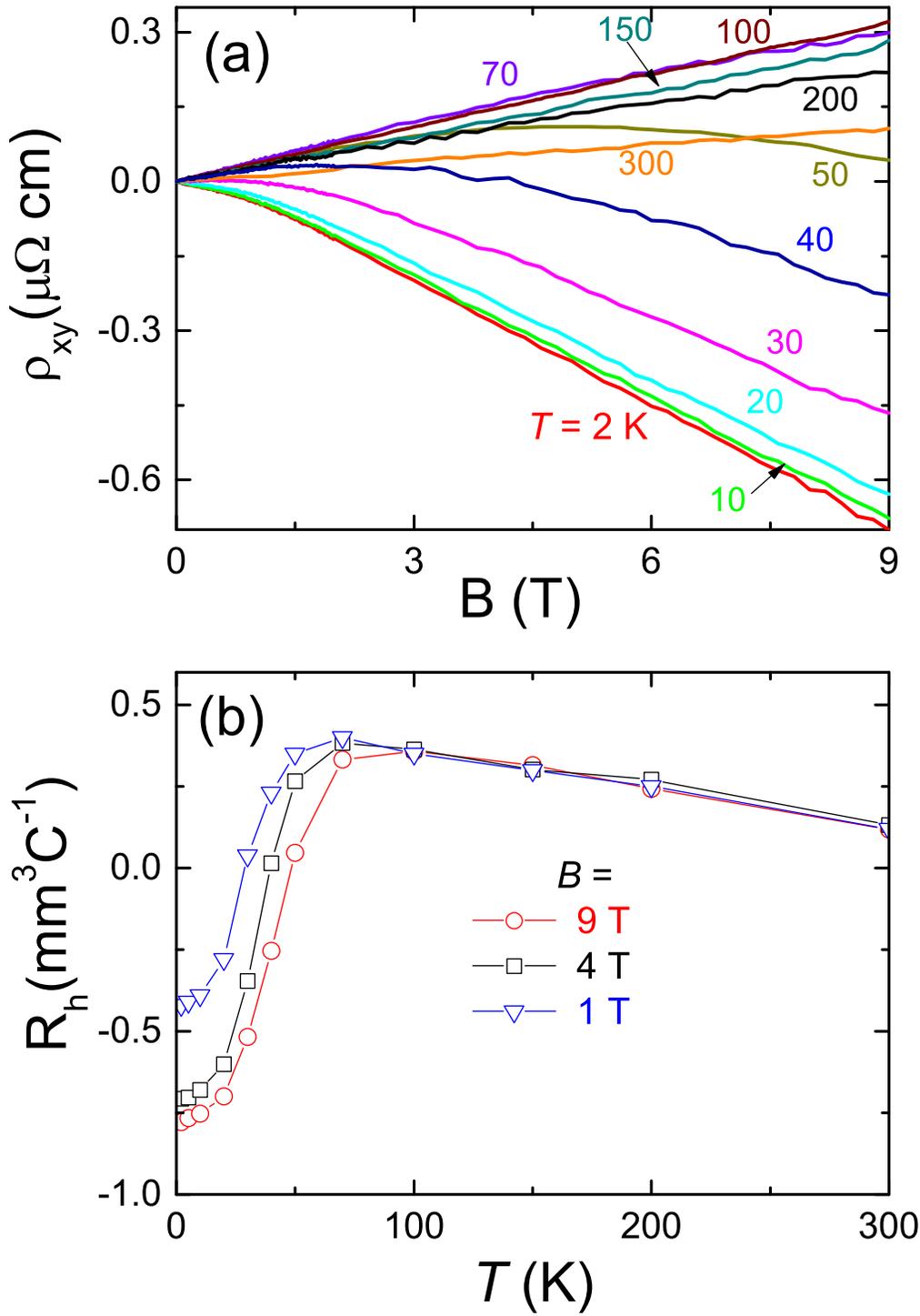}
\caption{{\bf Hall effect for MoAs$_2$.} (a) Magnetic field dependence of Hall resistivity at
several different temperatures up to 9 T. (b) Hall coefficient  vs. temperatures at 1 T , 4 T and 9 T. }
\end{figure}

\begin{figure}
\includegraphics[angle=0,width=16cm,clip]{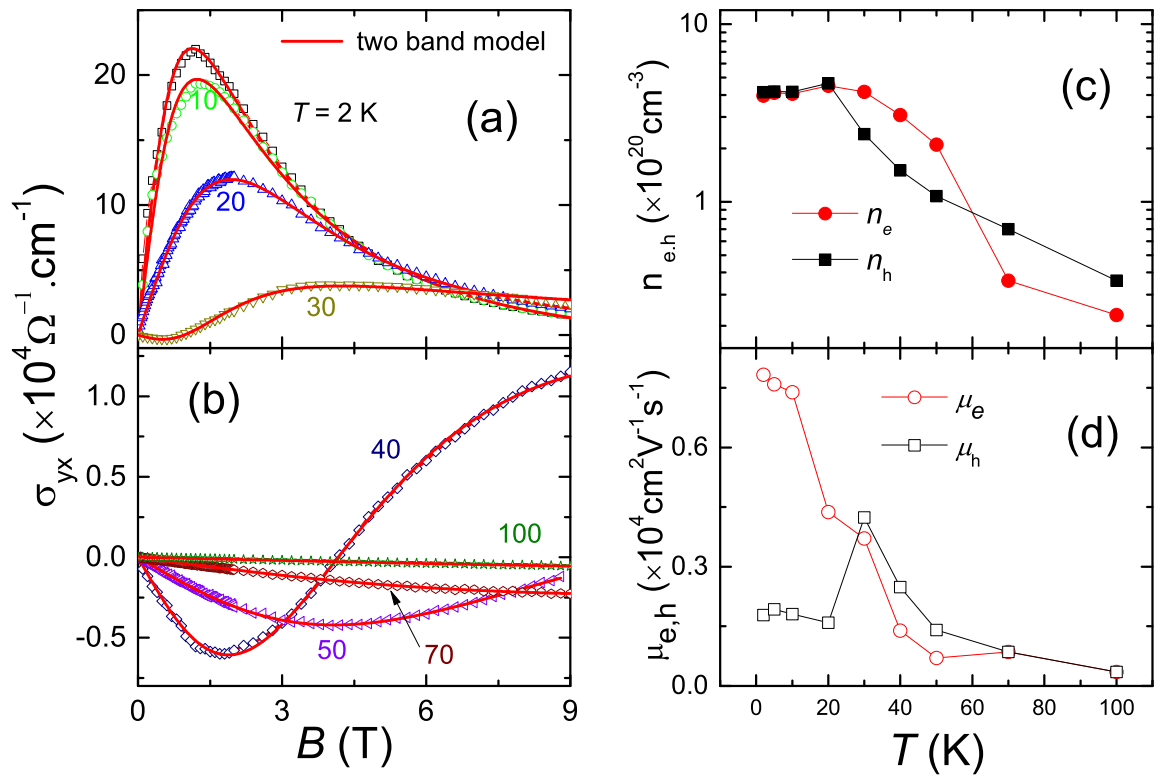}
\caption{{\bf Hall conductivity, the carrier density and mobility.} (a) and (b) Magnetic field dependence of Hall conductivity at several representative temperatures up to 9 T. The red solid lines are the fitting curves using two-band model.(c) Electrons and holes densities and (d) mobilities as a function of temperature below 100 K.}
\end{figure}


\begin{thebibliography}{10}
\bibitem{note} Detailed analysises for the topological nature of the band structure
and surface states will be given elsewhere.
\end{thebibliography}
\end{document}